\documentclass[twocolumn,prb,color,psfig,showpacs,showkeys]{revtex4}
\usepackage{graphicx}
\usepackage{epsfig}

\newcommand{\eps}{\varepsilon}

\begin{document}

\title{Nanoscale phase separation in $La_{0.7}Ca_{0.3}MnO_3$  films: \\ evidence for the texture driven optical anisotropy}

\author{A.S. Moskvin}
\author{E.V. Zenkov}
\email{eugene.zenkov@usu.ru}
\affiliation{Ural State University, 620083 Ekaterinburg, Russia}

\author{Yu.P. Sukhorukov}
\author{E.V. Mostovshchikova}
\author{N.N. Loshkareva}
\affiliation{Institute of Metal Physics, Ural Division of Russian Academy of Sciences, 620219 Ekaterinburg, Russia}

\author{A.R. Kaul}
\author{O.Yu. Gorbenko}
\affiliation{Moscow State University, 119899 Moscow, Russia}


\begin{abstract}
The IR optical absorption (0.1 eV $< \hbar\omega <$ 1.5 eV) in the
$La_{0.7}Ca_{0.3}MnO_3$ films on $LaAlO_3$ substrate exhibits the drastic
temperature evolution of the spectral weight evidencing the insulator to metal
transition. Single crystal films were found to reveal strong
linear dichroism with anomalous spectral oscillations and fairly weak
temperature dependence. Starting from the concept of phase separation, we
develop the effective medium model to account for these effects. The optical
anisotropy of the films is attributed to the texturization of the ellipsoidal
inclusions of the quasimetal phase caused by a mismatch of the film and
substrate and the  twin texture of the latter.
\end{abstract}

\maketitle

\section{Introduction}

The investigation of doped rare-earth manganites has a long history, starting from 1950th.
The intense researches of last decade, stimulated with the discovery of colossal
magnetoresistance  have enriched significantly our understanding of these systems.
However, recent studies argue its internal nature to be by far more complex as compared to
what is predicted by the conventional phase diagrams \cite{Urushibara, Schiffer}.
In particular, there is a growing experimental evidence \cite{phsep1, phsep2, Biswas} in favour of a generic
nanoscale inhomogeneity of manganites as well as many other strongly correlated transitional
metal oxides \cite{Pan, Uehara}.

The infrared (IR) optical studies \cite{Suh} of $La_{1-x}Ca_xMnO_3$ (0.1$< x <$
0.8) were one of the first to attest the phase separation in
manganites. The analysis of diffuse neutron scattering studies of spin
correlations in $La_{1-x}Ca_xMnO_3$ ($x\,<\,0.2$) single crystal \cite{Hennion}
confirms the existence of ferromagnetic (FM) inhomogeneities below $T_c$ in the
form of platelets with a mean diameter of about 16 \AA.

In the films, unlike as in the bulk samples, the phase-separated state can
persist at still higher doping probably due to the specific properties of the
film-substrate interface stabilizing the novel phase. Indeed, the atomic force
microscopy of $La_{0.67}Ca_{0.33}MnO_3$ films deposited on $LaAlO_3$ substrate
\cite{Biswas} yields a direct visualization of coexisting FM metallic and
charge-ordered insulating phases. The authors suggest the key factor governing
the occurence of this phase separation to be the mismatch of the film and the
substrate, resulting in nonuniform compressive strains of the lattice and
concomitant inhomogeneities in exchange and hopping parameters. The low strain
well conducting islands with mean size as large as 500 \AA, separated by high
strain insulating interfaces are clearly discernible on the micrographs of the
film.

In this paper we present the results of optical studies of strained $La_{0.7}Ca_{0.3}MnO_3$ films, that can be considered as another possible manifestation of intrinsic phase inhomogeneity  of these systems. Surpisingly enough, the IR absorption spectra display significant optical anisotropy and unusual oscillating frequency dependence of the dichroism, that cannot be associated with a certain electronic transitions. Our model calculations, based on the concept of strain-induced ordering of prolate quasi-metal droplets in parent insulator, agree quantitatively with experiment.

\section{Experimental}

Two $La_{0.7}Ca_{0.3}MnO_3$ ($LCMO$) manganite films, 60 nm and 300 nm in
thickness, were grown on single crystalline (001) oriented $LaAlO_3$ ($LAO$)
substrates. The details of films growth and sample characterization are given
elsewhere \cite{[1],[2]}. The X-rays diffraction patterns and high resolution
transmission electron microscopy confirm the epitaxial character of the films.
The rocking curve methods yields $FWHM=0.16^{\circ}$, that implies small
mosaicity and high quality orthorhombic $Pnma$ structure of the films. The
out-of-plane lattice constants $c$'s (3.870  \AA  (60 nm) and 3.872  \AA  (300
nm)) are close to the in-plane ones (3.863  \AA  (60 nm) and 3.862  \AA  (300
nm)) for both films, so that structural anisotropy is small. We note a small
amount of $Mn_3O_4$ impurity in the thick (300 nm) film, that does not alter
the perovskite structure of the sample and is observed in the form of nanoscale
embeddings in epitaxial matrix.

The optical experiments were performed in the range of $0.1<\hbar\omega<1.5$ eV,
$80<T<295$ K using automatic prismatic spectrometer. The absorption coefficient ($K$) was derived from the measured ratio of transmitted to
incident beams intensities as $K=(1/d)\ln \left((1-R)^2 I_0/I\right)$,
 $d$ and  $R$ being the film thickness and the reflectance, respectively.
 The grating polarizer was employed in IR region.
  The spectra were taken for the light wave {\bf E}-vector, adjusted along and
   normally to the direction of maximal absorption ($c$-axis), determined by
 the  rotation of the polarizer. As usual, the linear dichroism is defined as
\begin{equation}\label{diff}
 \Delta\,=\,\frac{K(E \parallel c)\,-\,K(E \perp c)}{K(E \parallel c)\,+\,K(E \perp c)},
\end{equation}
where $K$'s are the absorption coefficients for  appropriate polarizations.

\section{Results and discussion}

\subsection{Optical spectra}\label{3.1}

The IR optical response of doped manganites such as $La_{1-x}(Sr,Ca)_xMnO_3$
differs strongly from those of pure $LaMnO_3$ system. A
self-consistent description of the carge-transfer (CT) bands in $LaMnO_3$  \cite{[3]} shows the
multi-band structure of the CT optical response
  with the weak low-energy edge at $1.7$ eV, associated with forbidden
$t_{1g}(\pi)-e_{g}$ transition. These predictions are in a
good agreement with experimental spectra. A common feature of doped manganites
and related systems is revealed in an unconventional enhancement of IR spectral
weight, evidencing the appearance of free charge carriers contribution.

\begin{figure}
\begin{minipage}[b]{0.48\linewidth}
\includegraphics[width=\linewidth,angle=0]{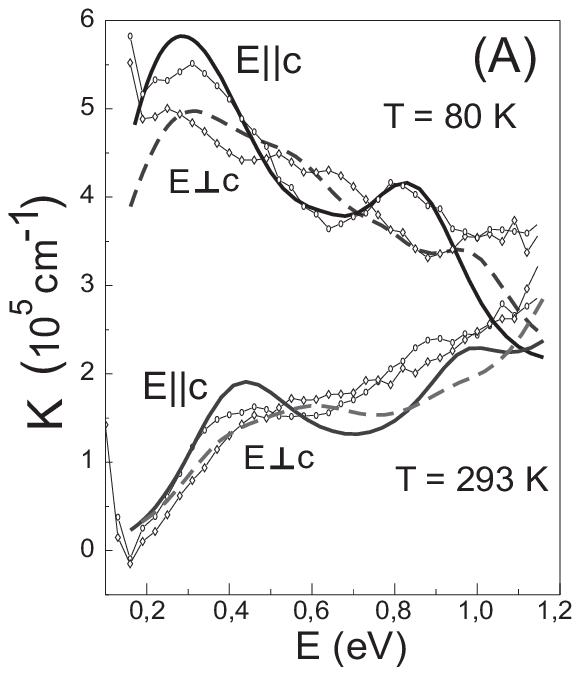}
 \end{minipage}\hfill
 \begin{minipage}[b]{0.48\linewidth}
\includegraphics[width=\linewidth,angle=0]{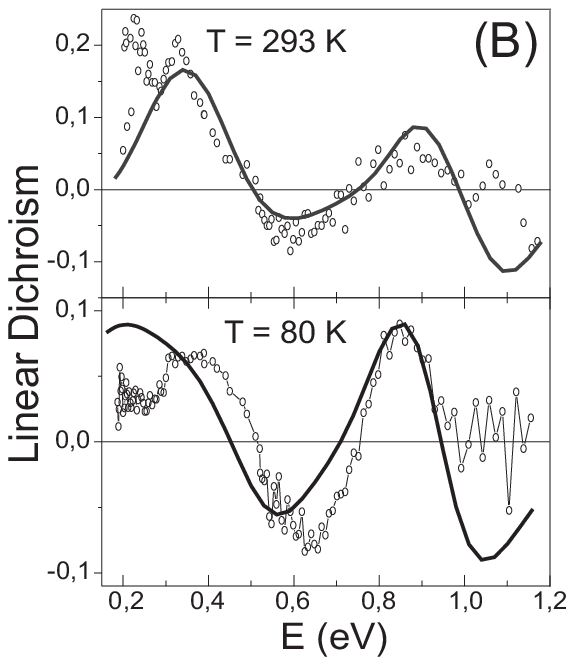}
 \end{minipage}
 \begin{minipage}[b]{0.48\linewidth}
\includegraphics[width=\linewidth,angle=0]{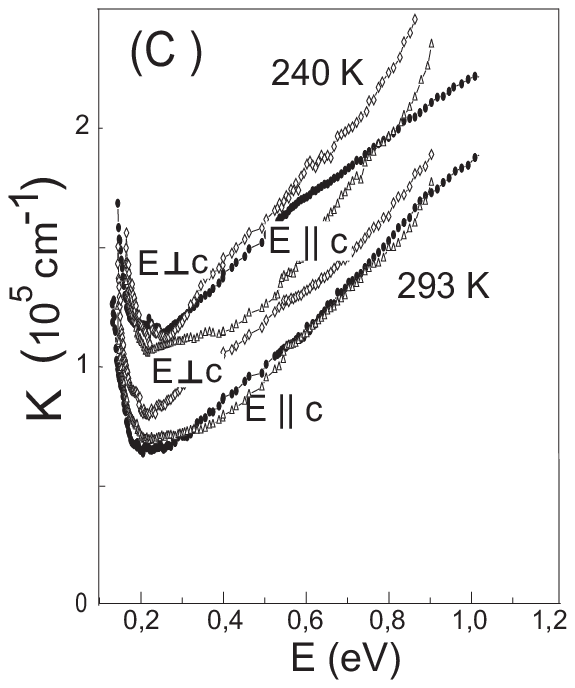}
 \end{minipage}\hfill
 \begin{minipage}[b]{0.48\linewidth}
\includegraphics[width=\linewidth,angle=0]{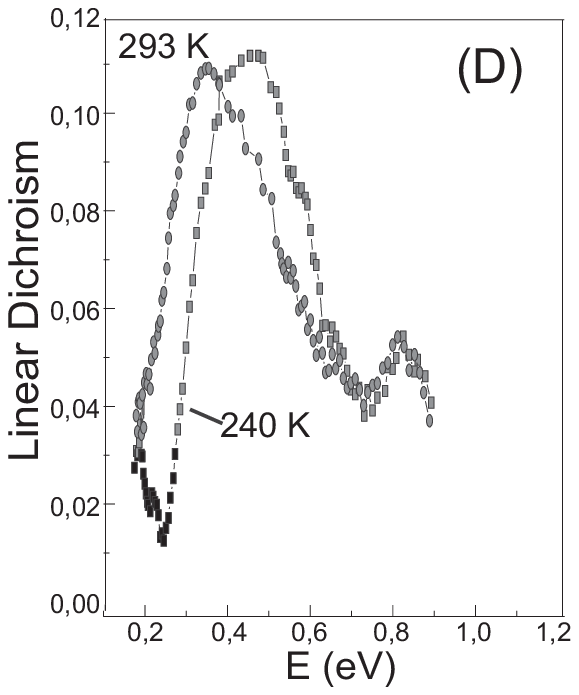}
 \end{minipage}
\caption{Top panel: Spectroscopic data for  thin (60 nm)
$La_{0.7}Ca_{0.3}MnO_3$ film at $T=80$ K and  $T=293$ K. (A) Absorption spectra
for two polarizations: experimental data are shown by dots,  curves are the
result of fitting in frames of effective medium theory (the unpolarized light
spectra are not shown). (B) Linear dichroism spectra: experiment (dots) and
effective medium theory (curves).
\\Bottom panel: Spectroscopic data for thick (300 nm) $La_{0.7}Ca_{0.3}MnO_3$ film
at $T=240$ K and  $T=293$ K. (C) Absorption spectra: unpolarized light
(circles),  {\bf E} $\parallel$ {\bf c} polarization (triangles),  {\bf E}
$\perp$ {\bf c} polarization (diamonds). (D) Linear dichroism spectra.}
\label{fig}
\end{figure}

The absorption spectra of thin (60 nm) film are shown in Fig.1A.
At $T\,=\,295$ K the unpolarized light IR spectrum displays an insulating gap-like
behavior. However, this behaviour is unusual, since the
absorption, without being metallic, remains rather large in entire spectral range.
The increase of absorption coefficient below
a dip near $0.2$ eV is related to phonon bands, starting at $\sim 0.09$ eV,

The lowering of  temperature to $80$ K, well below the Curie
temperature $T_c \approx 268$ K, entails a drastic enhancement of IR spectral
weight, straightforwardly evidencing the appearance of free charge carriers
contribution. Very large free carriers absorption of thick (300 nm) film at
$T=80$ K makes the transmission experiments difficult, so that in FM region the
measurements were performed only around $T_c$ (Fig.1C).We observed quite
similar behaviour also in self-doped $La_{0.83}MnO_3$ \cite{[4]} and
$La_{0.7}Sr_{0.3}MnO_3$ films \cite{[5]}. Comparative analysis of these spectra
with those of other doped manganites ($La_{1-x}(Sr,Ca)_xMnO_3$ \cite{Takenaka,
Jung}, $Nd_{1-x}Sr_xMnO_3$ \cite{LeeNoh}, etc.) allows us to conclude, that
here we deal with the seemingly universal physical behavior, common to a wide
family of doped manganites.  Earlier, we assigned  \cite{[4],2,3}
this unconventional optical response to nanoscopically inhomogeneous texture of
manganites which looks like a system of metallic droplets in insulating matrix.
The spectral features are believed to be governed mainly by the temperature
dependent metallic volume fraction and  geometrical resonances in a
nanoscopically inhomogeneous system. From theoretical standpoint, the
possibility of the phase separation
in manganites has been examined in pioneering work by Nagaev \cite{ferron}, developed
afterwards \cite{Khomskii}. It was shown, that the competition of hopping energy and
double exchange render the charge carriers segregation in nanoscopic FM conducting droplets (ferrons)
energetically favorable as compared to uniform spin-canted state.
Such a segregation is accompanied
with the gradual shift of the spectral weight from the absorption bands of
parent AF matrix to lower-energy excitations in the droplets of a novel phase.

The experiments in polarized light revealed the interesting features, observed
in IR absorption spectra of thin film both in FM and paramagnetic (PM) states,
viz. the oscillations about the unpolarized light spectrum (Fig.1A). The
spectra of thick films look differently (Fig.1C). Above $T_c$, the absorption coefficient
in {\bf  E} $\perp$ {\bf c} polarization exceeds in magnitude that in
unpolarized light, while in {\bf E} $\parallel$ {\bf c} polarization these
spectra  nearly coincide over the range $0.2$ to $0.9$ eV. In FM state (240
K), the {\bf E} $\perp$ {\bf c} and unpolarized spectra are indistinguishable
in $0.2-0.5$ eV range, but above $0.5$ eV the absorption coefficient in the {\bf E}
$\perp$ {\bf c} polarization shows more rapid increase. The {\bf E} $\parallel$
{\bf c} spectrum approaches the unpolarized one from below and intersects at
$0.8$ eV.

\begin{figure}
\includegraphics[width=2.7in, height=3.4in,angle=270]{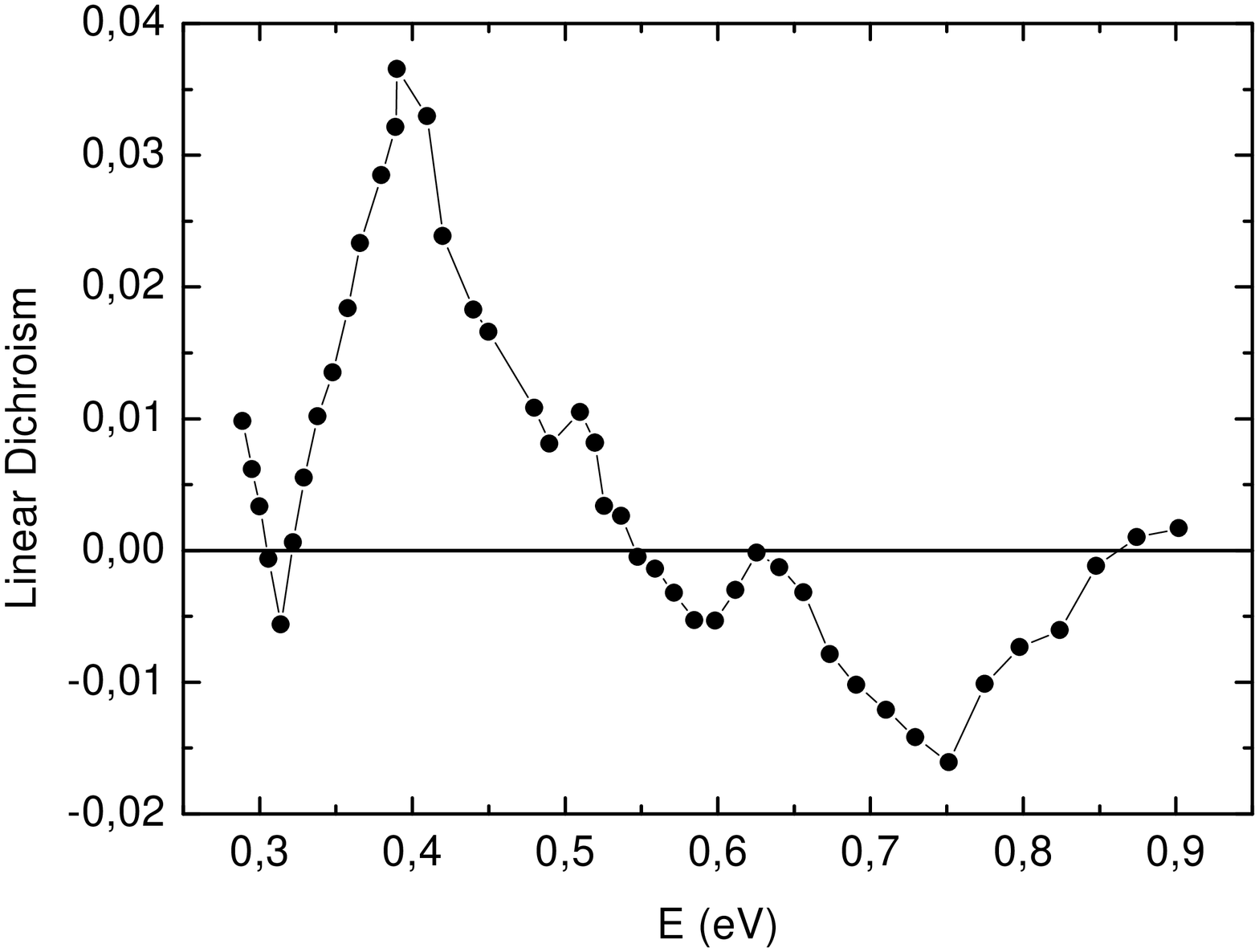}
\caption{The linear dichroism of the $LAO$ substrate.}
\label{fig2}
\end{figure}

The above peculiarities of polarized light absorption manifest itself in the
spectral oscillations of the linear dichroism of thin film with the amplitude
of the order of 0.3 about zero (Fig.1B). The dichroism  in thick film is
positive throughout the spectral range $0.2\div 0.9$ eV with
 two peaks, at $0.4$ eV  and  $0.8$ eV, which position is close to that
 observed in the  dichroism spectrum for thin film (Fig.1D).
 It should be noted, that upon cooling from $T=293$ K
 to $T=240$ K, the low-energy peak blue-shifts from $0.35$ to $0.45$ eV.

\subsection{Discussion}\label{3.2}

Nontrivial questions arise about the origin of the observed peculiarities of
optical  response, since no allowed electro-dipole transitions are known to
fall in the spectral range under consideration.
It is worthy of noting that the LAO substrate reveals itself the
optical anisotropy, most probably due to a twinning. However, its
absorption in the spectral range under examination is negligibly
small as compared with that observed for LCMO/LAO films. Hence,
small dichroism of substrate with its irregular spectral
dependence (see Fig. \ref{fig2}) can be neglected in our further
discussion.

While the magnitude of the absorption coefficient for the films varies by hundreds
of percents upon cooling from $T \sim 293^{\circ}$ K to $T=80^{\circ}$
K, the behaviour of dichroism remains practically unchanged both in PM and  FM
states, and its amplitude does not change significantly when turning from the
thin film to the thick one. Note, that the dichroism is observed in both films
and thus is not related to the $Mn_3O_4$ impurity effects.  However,
restricting ourselves to the conventional explanations, we may
overlook another possibilities, stemmed from the intrinsic nanoscale phase
inhomogeneity of doped manganites. In particular, we assume, that under the
mismatch induced nonuniform stresses, the droplets of the quasimetallic phase
may align to form somewhat like a nematic liquid crystal. Such a texture, reflecting the twin pattern of the substrate, results in a significant enhancement of the observed dichroism of the  $LCMO/LAO$ films. We show, that this
model, being the development of our general approach to manganites as
nanoscopically inhomogeneous systems  \cite{[4],2,3}, can provide a good
quantitative agreement with all experimental findings presented in the previous
section.

First, we would like to shortly overview the effective medium theory (EMT)
\cite{emt} which appears to be a powerful tool for the quantitative description
of the optical response of inhomogeneous systems.
In its simplest form, the EMT equation for effective dielectric function
$\eps_{eff}$ of the two-component inhomogeneous system reads as follows
\cite{emt}:
\begin{equation}\label{ema}
\int \limits_{V} dV
p(\eps)\,\sum_{i=1}^{3}\,\frac{\eps\,-\,\eps_{eff}}{\eps_{eff}\,+\,L_i\,(\eps\,-\,\eps_{eff})}
\,=0,
\end{equation}
where the integration runs over the volume of the sample,
$p(\eps)\,=\,p_1\,\delta(\eps(r)\,-\eps_1)\,+\,p_2\,\delta(\eps(r)\,-\eps_2)$.
We consider the two components of binary composite on equal footing with
$p_{1,2}$ being the volume fractions, $\eps_{1,2}$ and $L_{i}$  the dielectric
functions and the depolarization factors of the grains of its constituents,
respectively.

It is worth to emphasize that in effective medium approximation the volume of the droplet does not enter the calculations, but is accounted implicitly in the validity range of the theory, restricting the mean size of the droplet to be smaller than the wavelength. Hence, within the EMT, the volume fractions $p_{1,2}$ can only change through the number of the droplets rather than due to the variation of their sizes. In general, $p_{1,2}$ are
determined by thermodynamical conditions and depend on temperature, pressure,
and other external factors. The approach \cite{PShg}, we employed here to lend
more plausibility to this physically transparent EMT scheme, is to take into
account the natural difference between the core and the surface properties of
the inclusions using the standard expression for the polarizability of the
coated ellipsoid \cite{Bilboul}.

The spectra of nanoscopically disordered media display some specific features
due to geometric resonances \cite{emt}, that have no counterpart in homogeneous
systems. These arise as a result of resonant behaviour of local field
corrections to the polarizability of the granular composite and are governed to
a considerable extent by the shape of the grains. The frequency of geometric
resonance is then easily obtained as the one at which the polarizability of
small particle diverges. For the case of spherical metallic droplets embedded
in the insulating matrix with dielectric permittivity $\eps_d$ this leads to
the equation:
\begin{equation}
 \eps(\omega)_{part.}\,+\,2\,\eps_d\,=\,0,
\end{equation}
whence the resonance frequency is
\begin{equation}
  \omega_r\,=\,\frac{\omega_p}{\sqrt{1\,+\,2\,\varepsilon_d}},
\end{equation}
if the Drude's expression  with the plasma frequency $\omega_p$ is assumed for
the metallic permittivity and $\eps_d$ is constant. In the case of arbitrary
ellipsoid there are three different principal values of polarizability and the
latter formula generalizes to
\begin{equation}\label{ell}
\omega_r^i\,=\,\omega_p\,\sqrt{\frac{L_i}{\varepsilon_d\,-\,L_i\,(\varepsilon_d\,-\,1)}},
\ i\,=\,1,\,2,\,3,
\end{equation}
where $L_i$ are three shape-dependent depolarizaton factors. For the cubic
matrix it is naturally to assume  these ellipsoidal droplets to be oriented
along main symmetry directions like [111], [100].

Such a simple model of phase separation enables us to understand the nature of
linear dichroism observed in $LCMO/LAO$ films both above and below $T_c$ even
without any assumptions concerning the intrinsic electronic anisotropy of its
constituents. Indeed, for a bulk manganite crystal with cubic symmetry all
orientations of ellipsoidal droplets like [111] are  energetically equivalent
and equally distributed that restores the high symmetry of the system as a
whole and results in the optical isotropy of nanoscopically phase-separated
sample. However, any anisotropic noncubic perturbation like strain lifts the
orientational degeneracy of ellipsoidal droplets and gives rise to a certain
texture. When this texture is irradiated by
 a plane wave, the geometric resonances (Eq.\ref{ell}) will be excited selectively depending
 on the polarization. Thus, the phase-separated sample as a whole can exhibit sizable optical
 anisotropy in the spectral range of the geometric resonances. Namely such a
 situation is likely to occur in our samples due to the unavoidable
 mismatch of the film and the substrate that generally results in non-uniform strains, that can
 favor some ordering of the droplets to form a texture.
Since the film and the substrate have the same lattice structure, their
mismatch would in principle result in bulk contraction, changing the volume of
the unit cell. The manganite films grown on $LAO$ substrate are known
\cite{ThinFilm} to be under compressive stress, their lattice being contracted
in the plane and expanded in the normal direction. High energy cost of this
deformation makes more feasible the relaxation channel that implies the
nonuniform distribution of the metal volume fraction with the density enhanced
near the $LCMO/LAO$ interface. Indeed, the lesser  volume of the unit cell for
metallic phase as compared with insulating $LCMO$ matrix provides the
optimal relaxation of the mismatch. However, the large density of metallic
ellipsoidal droplets implies its ordering, or packing along one of four
equivalent directions in the plane of film. In our case this direction is
likely to be determined by the twin texture observed for $LAO$ substrate.
 In such a way we come to the
nucleation of the three-dimensional texture of metallic ellipsoidal droplets
resulting in the optical anisotropy of the film.
Our preliminary experimental examinations of the optical anisotropy of the twinned single-crystalline $La_{0.93}Ce_{0.07}MnO_{3}$ sample also argue in favour of this picture.
Note, that in all cases the resulting deformation of the lattice may be quite small to be
detected by conventional X-ray methods.

To check the validity of the hypothesis of the texture-driven dichroism in
$LCMO/LAO$ films we simply assumed the nanoparticles to be identically aligned
along the optical axis in the plane of the film, so that the amplitude of the
dichroism would be governed only by their shape anisotropy. To this end,
leaving only one of the terms in the sum (Eq. \ref{ema}), we are in the
position to calculate the eigenvalues of effective dielectric tensor and to
simulate the dichroism. For the model  to be more realistic we assume the
metallic-like droplet in the form of coated ellipsoid which  dielectric
function can be written as follows \cite{Bilboul}:
\begin{equation}
 \widehat{\eps}_i\,=\,\eps_{out}\,\frac{(\eps_{in}\,-\,\eps_{out})(f\,L_i^{out}\,-\,L_i^{in}\,-\,f)\,-\,\eps_{out}}{(\eps_{in}\,-\,\eps_{out})(f\,L_i^{out}\,-\,L_i^{in})\,-\,\eps_{out}},
\end{equation}
where $L$'s are the depolarization factors of inner ($in$) and outer ($out$)
confocal shells (core and coating, respectively), $f$ is the core to coating
volume ratio for this composite inclusion. Leaving aside the question of the
microscopic electronic structure of the droplet, we described its core and the
coating by the Drude formula:
      \begin{equation}\label{drude}
         \eps\,=\,\eps_0\,-\,\frac{\omega_p^2}{\omega\,(\omega\,+\,\mbox{i}\,\gamma)}
      \end{equation}
where $\omega_p$ is the plasma frequency, $\gamma$ is the damping parameter.
This approximation is reasonable and is more general that it may seem, since
the Drude form can be regarded as a limiting form of the universal expression
for dynamical conductivity in terms of memory function \cite{memory} with
$\omega_p$ and $\gamma$ to be effective parameters. Imaginary part of
dielectric function of $LaMnO_3$ was fitted to experiment \cite{Takenaka} by
the sum of three gaussian in a broad spectral range to ensure the validity of
its real part, derived via Kramers-Kronig transformation. Note, that the phonon
bands have been neglected throughout the calculations. The energies,
intensities and damping constants ($\omega$ and $\gamma$ being in units of eV)
of gaussians are: $(\omega,I,\Gamma)\,=\,\left((2.5, 2.6, 0.777), (5.0, 7.8,
0.929), (6.5, 2.866, 0.697)\right)$.

The main results of our calculations are shown in the Figs.1 A, B. The room
temperature spectra were fitted given the following parameters: the volume
fraction of quasi-metal phase $p=0.1$, $f=0.4$, $\omega_{p1}=3.0$ eV,
$\omega_{p2}=1.35$ eV, $\gamma_1=0.3$ eV, $\gamma_2=0.4$ eV, where the indices
$1,2$ stand for the core and the coating of the quasi-metal droplet,
respectively. The ratio of in-plane semiaxes of quasi-metal inclusions is set
equal to $\alpha=b/a=0.5$ and out-of-plane to major in-plane equal to
$\beta=c/a=0.42$. In principle, the quantum size effect in a small nonspherical
particle would bring about an anisotropic contribution to the relaxation rate
$\gamma$. For the sake of simplicity, we neglected this small effect in our
model calculations.

As follows from the experiment, the lowering of the temperature shifts the
phase equilibrium toward the ''metallization'', drastically expanding the
volume fraction of the metallic droplets. However, while the system becomes
more metallic, it may not be necessarily the case for an individual droplet
because of the noise, the random overlaps with neighbour droplets introduce in
its surrounding, so that the volume of the core relative to the fluctuating
edge region may even get smaller. To simulate the low-temperature spectra, we
modified the parameters ($\alpha$ and $\beta$ kept fixed) as follows: $p=0.55$,
$f=0.23$, $\omega_{p1}=2.4$ eV, $\omega_{p2}=1.12$ eV, $\gamma_1=0.3$ eV,
$\gamma_2=0.3$ eV. Thus, for a fixed doping, the temperature appears to be main
physical parameter, governing the metallic volume fraction and percolation. The
intrinsic "electronic" parameters like $\omega_{p}$ and $\gamma $, roughly
speaking, are  temperature independent with the accuracy of the order of
$10\div 20 \%$. The both results point toward the sound structure of our simple
model. Although it can hardly provide the excellent fit of experiment
throughout entire spectral range, it still captures the essential features of
the dichroism spectrum despite of a great body of obvious simplifying
assumptions. The two-peaked structure of the absorption coefficients, that
combines according to Eq. (\ref{diff}) to provide the oscillations of the
dichroism, results from the superposition of geometric (Mie's) resonances,
governed by a fine tuning of the parameters. It should be noted that the model
can yield rather complex behaviour of the spectra with multiple resonances. At
the same time it is worth noting, that similar calculations with simple
ellipsoidal particles fail to reproduce all the peculiarities of the absorption
and dichroism spectra. This means, that one should take care of the internal
structure of the nanoparticles when employing the EMT in realistic models.

In conclusion, we reported  the IR transmission measurements of the
$LCMO$ films grown on $LAO$ substrate. Upon cooling from
the room temperature to $T=80$ K, the absorption coefficient of thin film was found
to rise markedly.
The main finding of the present studies is unconventionally large nearly temperature
independent linear dichroism of the films, and its spectral oscillations, unexpected
in view of the good
structural perfection of the samples. We assert these features to be the manifestation of
the inhomogeneous phase-separated state of the films. The optical anisotropy of
the film is attributed to the texturization of the ellipsoidal nanoparticles of
the quasimetal phase caused by a mismatch of the film and  substrate and the
 twin texture of the latter. The simulation in frames of an effective
medium model provided a good description of experiment.
Despite this, at present we cannot exclude another possible scenarios of linear
dichroism in manganite films, in particular, the magnetic one. Indeed, our
model implies the intrinsic ferromagnetic ordering of quasimetallic droplets
with superparamagnetic behavior at $T>T_C$ and magnetic percolation below
$T_C$. Such a ferromagnetic droplet would manifest a magnetic linear dichroism
irrespective of its shape. However, for the film to reveal the linear dichroism
we need its relevant magnetic texture, or some kind of the ordering of the
droplets. One should note that similar to our model the quantitative
description of magnetic mechanism may be carried out in frames of effective
medium theory. Actually, the distinction of two mechanisms needs in further
studies.

In any case we see, that the analysis of the optical anisotropy provides the
effective tool for the examination of nanoscale texture of the film. More
generally, we suggest that the results of  present paper as well as a great
body of previous contributions \cite{Pan,Biswas,Hennion,[4]} have to enrich and
extend the conventional understanding of the optical response of manganites,
demonstrating the importance of specific effects of its intrinsic nanoscale
phase inhomogeneity. In particular, we may state that the specific properties
of dichroism in the spectral range 0.1 eV $< \hbar\omega <$ 1.5 eV support our
model \cite{[4],2,3} in which the appropriate optical response is governed
mainly by the geometrical resonances in nanoscale inhomogeneous system.

\begin{acknowledgments}
 The work was
supported by INTAS 01-0654, Federal Program (Contract No. 40.012.1.1. 1153-14/02),
grants RFBR No. 01-02-96404, No. 02-02-16429, RFMC No. E00-3.4-280, CRDF No.
REC-005 and UR 01.01.042.
\end{acknowledgments}

\end{document}